\documentclass[preprint,12pt]{elsarticle}
\usepackage{amssymb}

\journal{Chaos, Solitons and Fractals}

\begin{document}

\begin{frontmatter}

\title{Community structure in real-world networks from a non-parametrical synchronization-based dynamical approach}
\author{Abdelmalik Moujahid, Alicia d'Anjou, Blanca Cases}
\address{Department of Computer Science, University of the Basque Country UPV/EHU, 20018 San Sebastian, Spain}

\date{\today}

\begin{abstract}
 This work analyzes the problem of community structure in real-world networks based on the synchronization of nonidentical coupled chaotic R\"{o}ssler oscillators each one characterized by a defined natural frequency, and coupled according to a predefined network topology. The interaction scheme contemplates an uniformly increasing coupling force to simulate a society in which the association between the agents grows in time. To enhance the stability of the correlated states that could emerge from the synchronization process, we propose a parameterless mechanism that adapts the characteristic frequencies of coupled oscillators according to a dynamic connectivity matrix deduced from correlated data. We show that the characteristic frequency vector that results from the adaptation mechanism reveals the underlying community structure present in the network.

\end{abstract}


\end{frontmatter}

\maketitle

\section{Introduction}
\label{sec:1}

Community structure appears in many networked systems, including a variety of biological, social, technological, and information networks \cite{1}. A network can be represented as a graph G composed by a large number of highly interconnected units, and a community is characterized by a large number of edges connecting vertices within individual groups, with only low concentrations of edges between these groups \cite{2}.
The aim of community detection in graphs is to identify the modules and, possibly, their hierarchical organization, by only using the information encoded in the graph topology.
The problem has a long tradition and it has appeared in various forms in several disciplines \cite{3,4,5,6}. The general notion of community structure in complex networks was first pointed out in the physics literature by Girvan and Newman \cite{2}. During the last years many algorithms  have been proposed ranging from traditional methods (Graph partitioning, spectral clustering), modularity-based methods and synchronization-based dynamics algorithms \cite{7,8,9}. For a recent review about the problem of detecting communities in graphs see \cite{10}.

Dynamics algorithms based on synchronization have been studied by several authors \cite{11,12,13,14}, and have shown that the dynamical synchronization process in complex networks is a suitable method to unravel their different topological scales. Most of these works are based on the Kuramoto model \cite{15} or on its variants such as the opinion changing rate model \cite{16}, where the dynamics of each isolated oscillator is modeled in a one-dimensional state space.
However, as actual systems are always described by high-dimensional complex dynamics that involve many variables and parameters, it is of practical interest to know about community structure in complex networks in these circumstances. This issue was approached in \cite{18} considering N identical R\"{o}ssler  oscillators coupled according to the dynamics of the opinion changing rate model, making use of the edge betweenness load matrix which implies an additional computational cost.
On the other hand, these algorithms are usually dependent on initial conditions and frequencies distribution of coupled oscillators, and give rise to unstable solutions.

In this work we model the dynamics of a given complex network using an ensemble of nonidentical coupled chaotic R\"{o}ssler  systems, each one characterized by a defined natural frequency. Such a model is chosen because R\"{o}ssler  system has been fairly well studied, in particular, in the context of complex networks \cite{17,18,19,20,21}. Others oscillatory systems can also be considered.

Our proposed approach tackles the problem of community structure in networked systems focusing on two aspects, the first one is to consider an ensemble of interacting nonidentical chaotic R\"{o}ssler systems where the coupling force increases linearly with time simulating a society in which the interconnections between the agents grow in time. In fact, the increase of interactions between the members of a society helps to stabilize a plurality of different non-interacting clusters \cite{16}. The values of coupling strength range from zero, which corresponds to an initial uncorrelated state, to a maximum value given by the ratio $\lambda_{max}$/$ \lambda_2$, representing an intermediate state of highly interconnected units forming local synchronized clusters. $\lambda_2$ and $\lambda_{max}$ are the first non-zero and maximum eigenvalues of the Laplacian matrix respectively. The Laplacian matrix of a graph is given by $L=D-A$, where $D$ is the diagonal degree matrix and $A$ is the adjacency matrix of the graph G.
The connection between the spectral information of the Laplacian matrix and the hierarchical process of emergence of communities has been reported in \cite{11}.

The second aspect is related to the influence that neighbors have to change the natural frequency (or opinion) of a particular oscillator (or agent), and therefore the emergence of the modular structure found in the network where each module will correspond to oscillators with common average frequency.
Initially, when the coupling is zero, each oscillator evolves according to its natural frequency giving rise to uncorrelated states. As soon as the coupling is engaged new correlated states emerge corresponding to oscillators that will evolve at the same average frequency. We show that a dynamical adaptation of the characteristic frequencies of oscillators, according to the idea of confidence bound \cite{22}, improves the stability of these emergent correlated states, and gives rise to a final frequency vector that brings into relief the network community structure. To this end, a parameterless frequency adaptation approach is proposed.

The paper is organized as follows. In section 2, the dynamics of the R\"{o}ssler oscillators network is introduced. In section 3 we will report the proposed parameterless and enhanced adaptive approach used to change the characteristic frequencies of coupled oscillators.  In sections 4 and 5 we describe both the social and computer-generated networks considered to test our approach. Finally, the results of numerical simulations are drawn in section 6.

\section{Network dynamics}

The dynamics of a graph G of N nonidentical coupled oscillators can be described as follow:
\begin{eqnarray}
  \dot{\xi}_i(t) &=& F(\xi_i)+\sum_j K_{ij} A_{ij} (\xi_j-\xi_i),
\end{eqnarray}
where $i=1,...,N$. $\dot{\xi}_i(t) = F(\xi_i)$, $\xi_i\in\Re^n$ is the local dynamics of each autonomous oscillator. Each oscillator can be characterized by a defined natural frequency which is the long-time average of the phase velocity. The oscillators are coupled with the coupling coefficients $K_{ij}$ over a predefined topology characterized by the connectivity matrix $A_{ij}$.  $K_{ij}$ indicates the coupling strength between the $i$th and $j$th oscillators. If oscillators $i$ and $j$ are connected, they have a coupling strength $K_{ij}=K$; otherwise, the coupling strength is $K_{ij}=0$.

In this work, we study an ensemble of many interacting R\"{o}ssler systems coupled through the $y$ component:

\begin{equation}
\begin{array}{l}
  \dot{x}_i(t) = -\omega_i y_i-z_i,\\
  \dot{y}_i(t) = \omega_i x_i+ay_i+\frac{K}{k_i}\sum_j^N A_{ij}(y_j-y_i),\\
  \dot{z}_i(t) = b+(x_i-c)z_i.
\end{array}
\label{equ1}
\end{equation}

All R\"{o}ssler oscillators are nonidentical, and the parameter $\omega_i=\omega_0+\Delta \omega_i$ selected randomly from a uniform distribution corresponds to the natural frequency of the individual oscillator \cite{23}. $\Delta\omega_i$ is the frequency mismatch between neighboring chaotic oscillators.
We set $a=0.2$, $b=0.2$, $c=5.7$ and $\omega_0=0.9$ so as to ensure the individual oscillator generates chaotic dynamics with phase coherent attractor. $k_i$ is the degree of the $i$th oscillator, K is the coupling strength,  and $A_{ij}$ are the elements of the adjacency matrix of G ($A_{ij}$ are 0 if vertices $i$, $j$ are not connected and 1 otherwise).
The $y$-coupling would guarantee the stability of the synchronized state for a coupling force large enough \cite{24}.

\section{The adaptive mechanism of characteristic frequencies}

For a coupling force $K=0$, each oscillator evolves according to its own dynamics resulting in uncorrelated states space variables. As soon as the coupling strength increases, it may be that several oscillators synchronize and oscillate with a common average frequency, whereas their neighbors have their own, different, frequencies. To enhance the stability of these emergent regimes, we propose a parameter-free dynamical adaptation mechanism to change the characteristic frequencies of oscillators.

We assume that each oscillator repeatedly changes its characteristic frequency by the median frequency of their oscillators of confidence, i.e, oscillators whose dynamics are confined within the confidence manifold.
To define this confidence manifold, we proceed as follows: throughout the synchronized process we monitor the temporal evolution of the $y$ component of each oscillator during a predefined time interval. Then, we compute the similarity matrix $S$ between the resulting time series. This matrix will be based on normalized Pearson's coefficients, and calculated as follows: let $p_{ij}$ denotes the Pearson's coefficient between two arbitrary time series $y_i$ and $y_j$ corresponding to the temporal evolution of the $y$ component of the oscillators $i$ and $j$ respectively.
These Pearson's coefficients are transformed to similarities that belong to the interval [0, 1]: $$s_{ij}=(p_{ij}-min(p_{ij}))/(1-min(p_{ij}))$$
where $min(p_{ij})$ denotes the minimum $p_{ij}$ over the $S$'s entries. Based on this similarity matrix, for each oscillator $i$, we can compute its average similarity to all other oscillators. Let $m_i$ denote this average.

The matrix $S$ is converted to a dynamic connectivity matrix $A^*$ that will reflect the dynamic of real associations between the oscillators during the synchronization process. This matrix is constructed according to an adaptive neighborhood process as follows:
\[ A^*_{ij}= \left \{
\begin{array}{ll}
A_{ij} & \mbox{ if  }  s_{ij}>m_i  \mbox{ or  } s_{ij}>m_j \\
0 &  \mbox{otherwise } \end{array} \right.  \]

Finally, for each oscillator $i$, we define its neighbors of confidence,  $\nu (i)$, as the set of adjacent oscillators according to $A^*$. So, the frequency of the oscillator $i$ is replaced by the median frequency of oscillators forming the confidence neighbors. Using the median instead of the mean frequency tends to be more robust to outliers, and gives rise to a frequency vector with a clear clustering structure.

This mechanism of frequencies adaptation is initiated once the coupled oscillators reach a certain degree of synchronization characterized by high values of average similarity $m_i$. We first consider a phase in which the coupled oscillators are forced to synchronize through the increase of the coupling strength and according to the underlying network topology. When the coupling strength $K$ reaches the appropriate value  \cite{20} for which complete synchronization is not established, but others synchronization features could emerge,  we keep it fixed.  Then, we began the mechanism of frequencies adaptation as described above. This mechanism takes place every few time units (or adaptation step) during a specific time interval. Both the simulation time and the adaptation step determine the length of the time series used to compute the similarity matrix. The number of times we adapt the oscillator frequencies  is given by the ratio between the time interval and the adaptation step, and should be limited to avoid a homogeneous frequency vector.

The adaptation procedure gives rise to a one-dimensional frequency vector with some clustering structure able to detect the underlying modules present in a given network. Each adaptation step a new frequency vector is obtained, and then new community subdivisions will emerge. So to determine the optimal number of modules or clusters we have adopted the criterion of maximum modularity \cite{25, 26}. The {\em modularity} is defined as  the fraction of links within communities minus the expected fraction of such links in a random network. This measure provides a way to determine if a certain description of the graph in terms of communities is more or less accurate. High values of modularity should indicate good partitions with many more internal connections than expected at random. This criterion has been adopted for two reasons: is the most widely used in the literature,  and the algorithms chosen to compare our methods adopt the same criterion. Although recently, it has been shown that the modularity maximization could exhibit extreme degeneracies admitting an exponential number of distinct high-modularity solutions depending on both the size of the network and on the number of modules it contains \cite{27}. In this work, the existence of possible degenerate solutions is limited since we have considered small real-world networks with few number of modules.

\section{Real-world networks}

In this paper, we have considered four real networks from the social science literature which have grown into standard tests for community structure detection algorithms. These networks are:
\begin{itemize}
 \item[-] The karate club network \cite{28} which shows the partition of friendships between the members of a karate club at US university in the 1970s. It consists of 34 members, and after the observation and construction of the network, the club in question splits in two as a result of an internal dispute.

 \item[-] The  network of bottlenose dolphins \cite{29} that was constructed from observation of a community of 62 bottlenose dolphins, where ties between dolphin pairs were established by observation of statistically significant frequent association. Bottlenose dolphins communities have been described as a fission-fusion societies and therefore individuals (or agents) can make decisions to join or leave a group. The community structure of this network reveals two major communities.  The first one is formed by 21 dolphins, while the second community which further splits into three sub-communities groups all other dolphins.

 \item[-] The jazz bands network \cite{30} which consists of 198 bands performed between 1912 and 1942. Two bands are connected if they have a musician in common. The bands network highlights the existence of two large communities, where the largest community also branches into two sub-communities.

 \item[-] The network of American football games between Division IA colleges during regular season Fall 2000. Nodes in the network represent teams and edges represent games between teams \cite{2}. This network is characterized by a known community structure in which nodes are divided into conferences containing around 8-12 teams each. Games are more frequent between members of the same conference than between members of different conferences.

\end{itemize}

\section{Tests on computer-generated networks}

Moreover, we have tested the performance of our method on computer-generated graphs with a known community structure. We have adopted the early class of the standard benchmark graphs of Girvan and Newman introduced in \cite{31}. Theses graphs are generated with $N=128$ nodes, split into four communities of 32 nodes each. Links between nodes belonging to same community are drawn with probability $p_{in}$, while a pair which belong to different community are joined with probability $p_{out}$. The value of $p_{out}$ is chosen so that the average number of inter-community edges per node, i.e, $z_{out}$, can be controlled. Each node have on average $z_{in}$ edges to nodes in the same community and $z_{out}$ edges to nodes in other communities, maintaining a total average node degree $k=z_{in}+z_{out}=16$.

We consider values of $z_{out}$ ranging from five, corresponding to clear community structure ($z_{out}\ll z_{in}$), to 10 which describes a network with bad defined structure ($z_{out} \gg z_{in}$). Since the "real" community structure is well known for these trial networks, we can validate our method computing the fraction of correctly identified nodes.

\section{Numerical results}

In this section, we integrate the system of Eq. [2] over the real-world networks described in the previous section. Each network is characterized by a connectivity matrix $A_{ij}$, and the initial characteristic frequencies $\omega_i$ which distinguish each oscillator. Values of  $\omega_i$ are randomly chosen from a uniform distribution with a frequency mismatch $\Delta\omega_i=0.2$, so that the R\"{o}ssler oscillators will evolve in a chaotic phase-coherent regime.

Numerical results have been performed simulating the system over $75$ time units.
We divided the simulation process into two phases. During the first stage ($time<20$), every one time units, the coupling force is increased linearly  from an initial value $K=0$ to a value not large enough to induce a regime close to complete synchronization. The values achieved are $K_Z=1.50$, $K_D=1.80$, $K_J=2.13$ and $K_F=0.9793$ for the Zachary, Dolphins, Jazz and Football networks respectively. In this stage, the coupled oscillators are forced to synchronize mutually according to the network topology and the interaction strength.

\begin{figure}[ht]
\begin{center}
\includegraphics[width=1\textwidth]{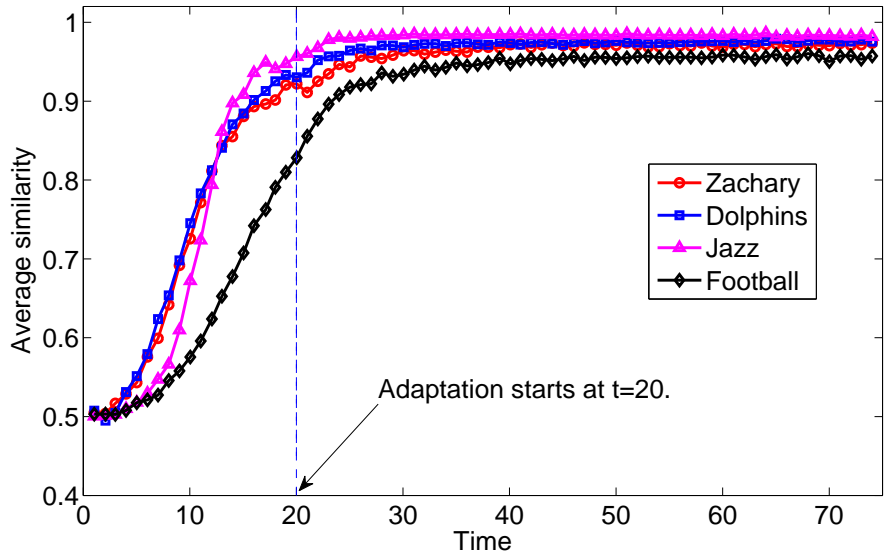}
\end{center}

\caption{The average similarity of the oscillator networks versus time performed over 100 runs of the algorithm.}
\label{figure1}
\end{figure}

In the second phase ($time>=20$), we keep $K$ fixed and then we began the process of frequencies adaptation.
As for the coupling force, every one time units, we adjust the oscillator frequencies according to the adaptation mechanism and we register values of average similarity $\langle m_i \rangle$ of each oscillator to characterize the temporal coherence of the network, and values of modularity $Q$ to quantify the goodness of a partition of the network into communities.
To compute $Q$, the nodes in the network are clustered according to the resulting one-dimensional frequency vector. For all the networks considered in this work, this vector has shown a clear community structure corresponding to oscillators that evolve at the same average frequency. To determine the optimal number of communities, we have computed the modularity for  a division of a network into different number of communities ranging from two to eight communities. The values of modularity achieved by our approach have been compared  with those obtained by the  widely used betweenness-based algorithm of Girvan and Newman \cite{2}.

In Fig.1 we report the global temporal coherence of the whole network $[\langle m_i \rangle]$, where $\langle . \rangle$ stands for the average of all oscillators and $[.]$ denotes averaging over 100 different system simulations.

\begin{figure}[ht]
\begin{center}
\includegraphics[width=1\textwidth]{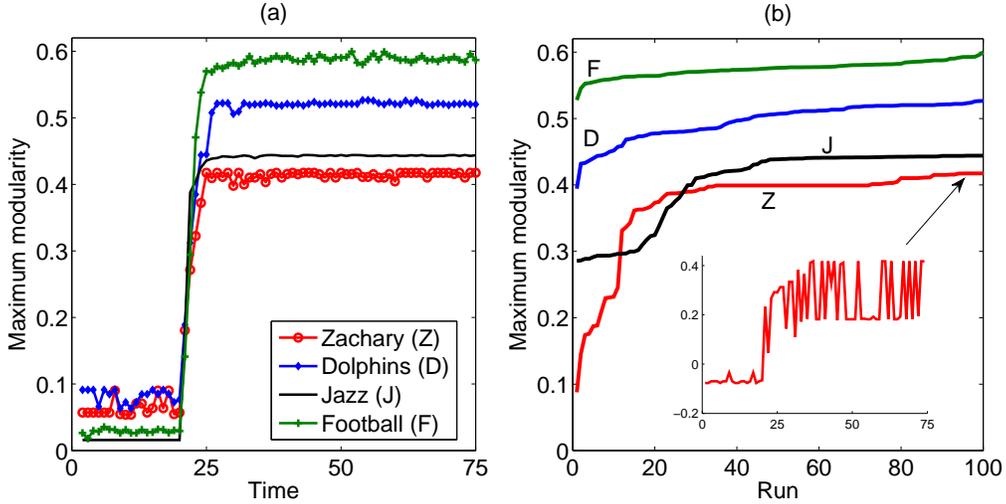}
\end{center}
\caption{(a) Time evolution of the maximum modularity $max(Q)$ achieved over a set of 100 different runs of the algorithm. (b) Maximum values of modularity corresponding to each run and reflecting the variability over the 100 different realizations. For Zachary and Dolphins the maximum modularity corresponds to a division into four communities. For the Jazz network the optimal division reveals three communities, while for Football network the optimal modularity is achieved for a partition into eight communities. In the inset of panel (b) temporal modularity of the Zachary network corresponding to one of the optimal runs.}
\label{figure2}
\end{figure}

As it can be appreciated, during the first stage, the network coherence shows an increasing pattern due to the gradual increase of the coupling force. It is important to stress that, in this region of the coupling strength, agents in each final group tend to maintain a synchronized regime despite their different characteristic frequencies. This is due to the strong influence the community structure present in the network which affects notably the emergence of synchronized groups.
After the adaptation mechanism takes place, we appreciate a slight increase of the global coherence of the network that we associate with the emergence of synchronized groups oscillating at the same average frequency. In this phase, as it can be seen in Fig 2 (a), we observe that the effect of the adaptation is immediate causing a sharp increase in the values of the temporal modularity $ Q(t)$. In fact, Fig. 2(a) reports the maximum values of modularity $Q_{max}$ achieved over a set of 100 different executions versus time. These values correspond to a division into four communities for the Zachary and Dolphins networks, while for the Jazz and Football networks the maximum values are achieved for a partition into three and eight communities respectively. All these subdivisions appear to show stable patterns with almost the same values of $Q_{max}$ in the case of the Dolphins and Jazz networks, and oscillating values with small deviations for the Zachary and football networks. On the other hand, according to the variability of the maximum values of modularity over the 100 runs of the algorithm (Fig. 2(b)), we can conclude that, in general, with few runs of the algorithm one can reach the maximum modularity values reported in Table 1. Data reported in Fig. 2(b) corresponds to sorted values of maximum modularity over the set of 100 runs of the algorithm. Indeed, we can appreciate that for the Zachary network  83\% of time, $Q_{max}> 0.363$ which is the value achieved by the edge-betweenness algorithm for a partition into four communities. For the Dolphins network, $Q_{max}$ is greater to the value $0.458$ associated with a partition into four communities by the edge-betweenness algorithm with a percentage of 88\%. The inset of Fig. 2(b) shows for the Zachary network an example of time evolution of modularity corresponding to one of the optimal runs.

\begin{table*}
\begin{center}
{\small
\begin{tabular}{|l|l|l|l|}
\hline\hline
Network  &  $Q_N$ ($N_C$)          &  $Q_{max}$($N_C$)       & size  \\ \hline
Zachary  &  0.4090 {\bf (4)}; 0.4010 (5) &  0.4174 ({\bf 4}); 0.4143 (5) & [13,11,5,5]  \\
Dolphins &  0.4580 (4); 0.5190 {\bf (5)} &  0.5220 {\bf (4)};0.5184  (5) & [20,12,9,21]  \\
Jazz     &  0.4379 (4); 0.4452 ({\bf 5)} &  0.4437 (4); 0.4436 (5) &    \\
         &                               &  0.4440 {\bf (3)}       & [61,75,62]  \\
Football &  0.5470 (6); 0.5980 {\bf (8)} &  0.5746  (6); 0.5901 {\bf(8)} & [20,10,15,25,12,11,9,13]  \\
 \hline  \hline

\end{tabular}}
\caption{Comparison of modularities for the network division found using our approach and betweenness-based algorithm of Girvan and Newman [Girvan et al. 2002]. $N_C$ is the number of communities found. $Q_N$ is the modularity achieved by the algorithm of Girvan and Newman.  $Q_{max}$ is the maximum modularity corresponding to the optimal configuration. Also included is the size of communities corresponding to the optimal partition. The number of communities with maximum values of Q have been highlighted.}
\end{center}
\end{table*}

In Table 1 we report values of modularity corresponding to the optimal partitions achieved taking our approach compared to those obtained by the betweenness-based algorithm of Girvan and Newman \cite{2}. Depending on the number of communities found using the two approaches, the difference in maximum modularity is about 5.5$\%$ for small number of communities and 1$\%$ when the number of communities is somewhat higher.

In the following we present, for each of the considered networks, the numerical outcomes illustrating the clustering of the characteristic frequency vector resulting from the adaptation process.

\bigskip

\subsection{The Zachary network}

In Fig. 3(a) we report values of $\omega_i$ versus node index in the Zachary network for a partition into two communities in accord with the natural division originally observed in \cite{28}.
The maximum modularity value achieved by our approach for this partition is $Q^Z=0.3718$. As it can be appreciated, $\omega_i$ values are visibly split into two main groups.
Nodes 3 and 10 are assigned to the erroneous community and display values of $\omega_i$ slightly lower than the average frequency $\langle\omega \rangle^1=0.9486$  associated with the community to which they belong.
Fig. 3(b) shows values of $\omega_i$ sorted in ascending order (circle marker) contrasted with the average frequency of each community (dashed line).

\begin{figure}[ht]
\begin{center}
\includegraphics[width=1\textwidth]{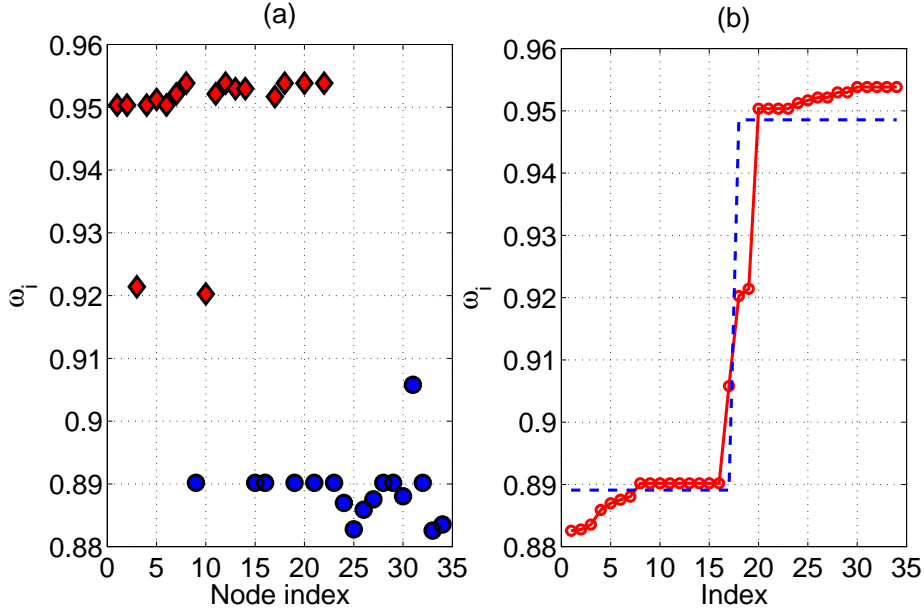}
\end{center}
\caption{The Zachary network. (a) The scatter plot illustrating the clustering of the characteristic frequencies of oscillators. (b) The sorted values of frequency of oscillators (circle marker) according to a division into two clusters achieved by our approach. In dashed line is represented the average frequency of each community.}
\label{figure3}
\end{figure}

On the other hand, the maximum modularity achieved by our approach is reached for a division of the Zachary network into four communities with a value $Q_{max}^Z=0.4174$ which is 14\% higher than the modularity obtained by the betweenness-based algorithm for the same partition. The resulting partition represents the division of the two initial communities into two sub-communities each one. The first community consists of 13 nodes oscillating at the average frequency $\langle\omega \rangle^1=0.8870$, while the second one with 11 nodes is characterized by the average frequency $\langle\omega \rangle^2=0.8949$. The last two communities both with five nodes each are identified by the average frequencies  $\langle\omega \rangle^3=0.8958$ and $\langle\omega \rangle^4=0.8976$ respectively.

\bigskip

\subsection{The Dolphins network}

The natural division of the Dolphins network originally observed in \cite{29} shows two major communities where one of them is further divided into three sub-communities. This partition has been detected using our adaptation mechanism leading to a frequency vector whose values $\omega_i$ versus node index are depicted in Fig. 4(a-b).
As it can seen, we can associate the first community to oscillators whose average frequency is about $\langle\omega \rangle^1=0.8768$. The number of nodes in this community is 21. All other oscillators are part of the second community which it could be divided into three sub-communities oscillating at the average frequencies  $\langle\omega \rangle^2=0.9497 $,  $\langle\omega\rangle^3=0.9617$  and  $\langle\omega \rangle^4=0.9433$  respectively. The distribution of nodes in these three sub-communities is as follow: 20 oscillators move at the average frequency $\langle\omega \rangle^2$, 12 at $\langle\omega \rangle^3$ and 9 at $\langle\omega \rangle^4$.
The modularity corresponding to this division has a value $Q^D=0.5220$ which is about $13\%$ higher than the value achieved by the betweenness-based algorithm.

\begin{figure}[ht]
\begin{center}
\includegraphics[width=1\textwidth]{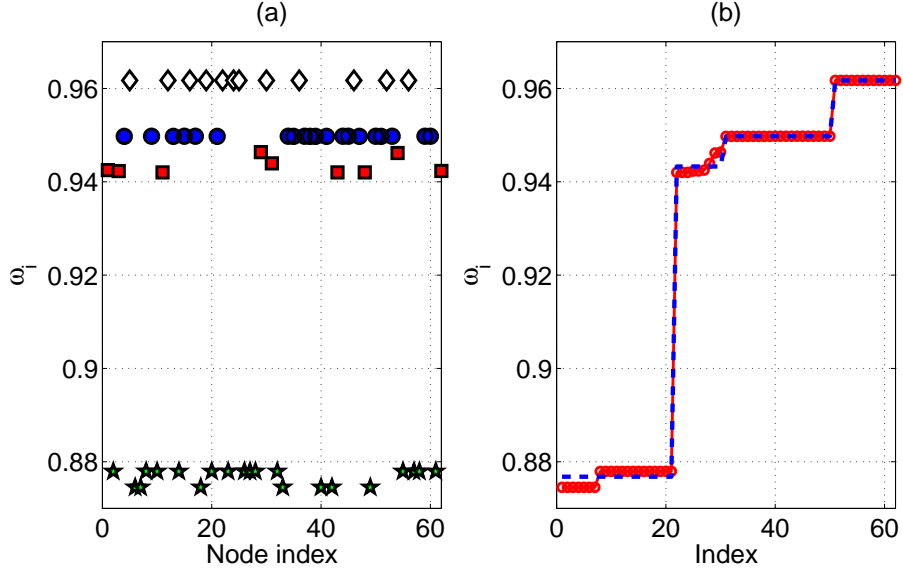}
\end{center}
\caption{The Dolphins network. (a) The scatter plot illustrating the clustering of the characteristic frequencies of oscillators. (b) The sorted values of frequency of oscillators (circle marker) according to a division into four clusters achieved by our approach. In dashed line is represented the average frequency of each community.}
\label{figure4}
\end{figure}

\bigskip

\subsection{The Jazz network}

In Fig. 5(a) we report values of $\omega_i$ versus node index in the Jazz network for a partition into  three communities in accord with the natural division originally observed in \cite{30}. The maximum modularity value achieved by our approach for this partition is $Q^J=0.444$.
As it can be appreciated, the three communities are clearly identified by the split of the frequency vector. The first one characterized by the average frequency $\langle\omega \rangle^1=0.8765$ is made up of 61 nodes. The second community with 75 nodes can be recognized by the characteristic average frequency $\langle\omega \rangle^2=0.9232$. In this case, we observe two nodes (nodes 17 and 74) that can be considered as outliers since they oscillate at a frequency much higher than the frequency characterizing this community. Finally, the third community is formed by 62 nodes oscillating at the average frequency $\langle\omega \rangle^3=0.9705$.

\begin{figure}[ht]
\begin{center}
\includegraphics[width=1\textwidth]{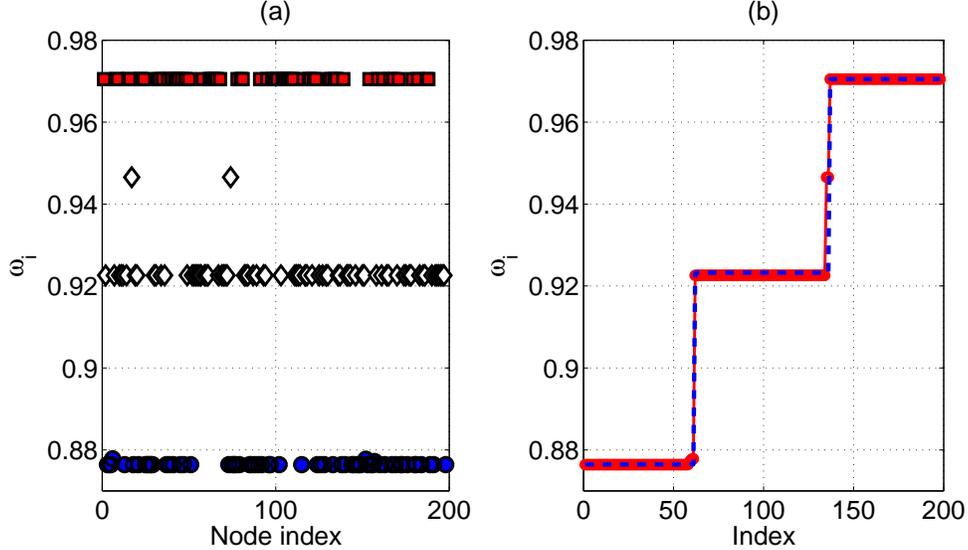}
\end{center}
\caption{The Jazz network. (a) The scatter plot illustrating the clustering of the characteristic frequencies of oscillators. (b) The sorted values of frequency of oscillators (circle marker)  according to a division into three clusters achieved by our approach. In dashed line is represented the average frequency of each community.}
\label{figure5}
\end{figure}

\bigskip

\subsection{The Football network}

Finally, as for the previous networks, the Fig. 6(a-b) report values of $\omega_i$ versus node index in the Football network for a partition into eight communities. One can clearly observe how the adaptive mechanism clusterizes the characteristic frequency vector resulting in oscillators that evolve at the same average frequency. The maximum modularity value achieved by our approach for this partition is $Q^F=0.5901$ which agree with the value achieved by the betweenness-based algorithm. This partition in eight communities is characterized by the following average frequencies: $\langle\omega \rangle^1=0.8428$, $\langle\omega \rangle^2=0.8615$, $\langle\omega \rangle^3=0.8767$, $\langle\omega \rangle^4=0.8949$, $\langle\omega \rangle^5=0.9244$, $\langle\omega \rangle^6=0.9454$,  $\langle\omega \rangle^7=0.9490$ and $\langle\omega \rangle^8=0.9508$. And corresponds respectively to clusters with 12, 20, 25, 13, 9, 15, 10 and 11 nodes each.
Further clustering of some identified communities, gives away that the oscillators with the average frequencies $\langle\omega \rangle^2$ (square marker) and $\langle\omega \rangle^3$ (circle marker) could be further divided into two sub-communities each one giving rise to a partition into ten communities.

\begin{figure}[ht]
\begin{center}
\includegraphics[width=1\textwidth]{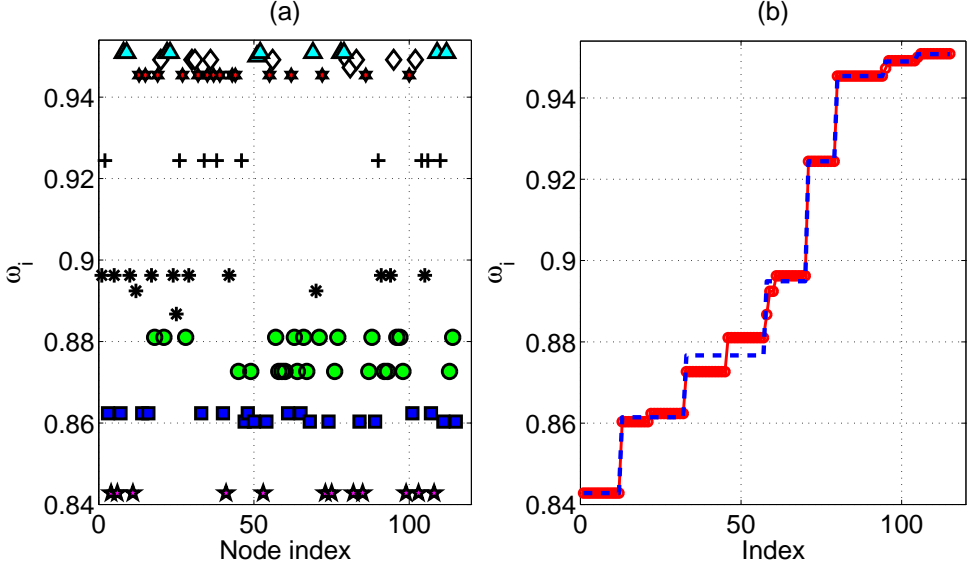}
\end{center}
\caption{The football network. (a) The scatter plot illustrating the clustering of the characteristic frequencies of oscillators. (b) The sorted values of frequency of oscillators (circle marker)  according to a division into eight clusters achieved by our approach. In dashed line is represented the average frequency of each community.}
\label{figure6}
\end{figure}

\bigskip

\subsection{The ad hoc trial networks}

To conclude, we report experiment results carried out with ad hoc trial networks as described in section 5. Figure 7 shows values of the fraction of correctly identified nodes, averaged over twenty-five different realizations of the computer generated networks, as a function of the average number of inter-community edges $z_{out}$. We have also registered values of the modularity $Q$ achieved by our approach (see inset of Fig. 7). These results have been compared with those achieved by the synchronization-based dynamical clustering algorithm (OCR-HK) of \cite{12}. The improvement in the performance of our approach is clearly appreciated as soon as  $z_{out}$ exceeds the value of six inter-modular edges per node (see Fig.1 in Ref. \cite{12}). The sensitivity of the OCR-HK algorithm starts to decay above $z_{out}\geq 6$, while for our method this decrease starts for $z_{out}> 7$. On the other hand, when the networks have a clear community structure ($z_{out}<6$) the performance of our method is complete achieving 100\% of prediction correctly. For this value, the OCR-HK identify  about 90\% of the nodes correctly, while our method is able to correctly classify 94\% of the nodes. For values of $Z_{out}>7$, both algorithms perform the same fraction of correctly classified nodes. It should be stressed that for values of $Z_{out}=8$ only few algorithms are still able to identify over 80\% of the nodes correctly \cite{32}, but with a hight computational cost.

\begin{figure}[ht]
\begin{center}
\includegraphics[width=1\textwidth]{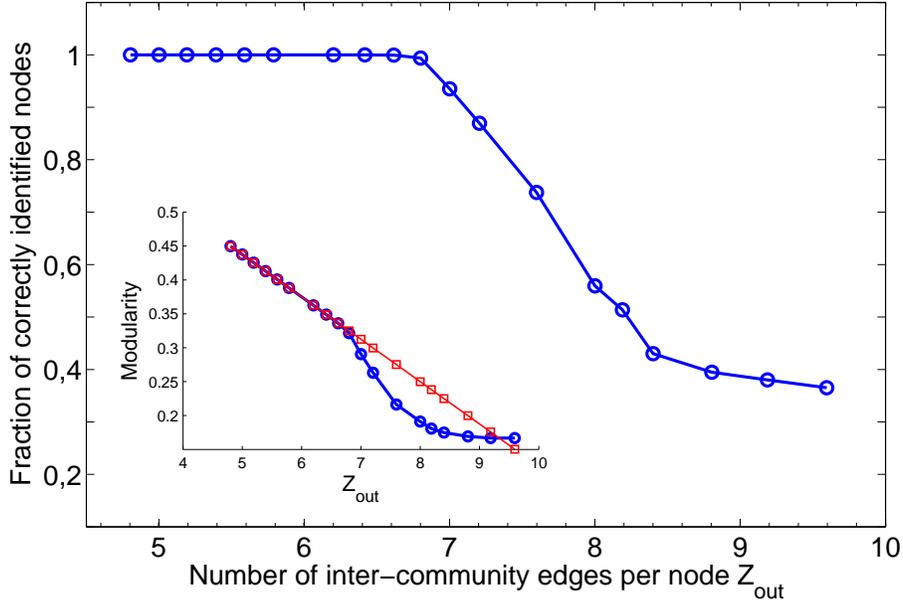}
\end{center}
\caption{Fraction of the correctly identified nodes versus the number of inter-community edges per node $Z_{out}$ averaged over 25 graph realizations. In the inset we report values of modularity $Q$ achieved by our approach (circle marker) and those corresponding to the real partitions (square markers) versus $Z_{out}$.}
\label{figure7}
\end{figure}

\bigskip

\section{Conclusions}
This paper considers the problem of community structure in complex real-world networks based on the synchronization of nonidentical chaotic oscillators. We have discussed a parameterless approach that adapts dynamically the characteristic frequencies of coupled oscillators to enhance the stability of the emergent synchronized groups according to a dynamic connectivity matrix. This adaptation mechanism gives rise to a frequency vector able to reveal the community structure presents in a given network. The proposed method has been tested both in real-world and ad hoc trial networks. The goodness of a given division of the network into communities has been quantified using the criterion of maximum modularity. The obtained values of modularity have been compared with those achieved by the betweenness-based algorithm of Girvan and Newman. The results are promising and fall within the scope of the results published in the literature for small networks with few number of communities. It should be stressed that the approach proposed in this work was inspired by the synchronization-based dynamical clustering algorithm (OCR-HK) of \cite{12} which scales as $O(N^2)$ tacking into account that it requires to compute the edge-betweenness of the network. So, in the worst case, our algorithm  will scale as $O(N^2)$ too, which is a reasonable computational cost to address larger networks.

\bigskip

\section{Acknowledgment}
The authors are very grateful to the editor and reviewers for their valuable comments and suggestions to improve the
presentation of this paper. Also, we would like to thank all those researchers that compiled and made public data sets and algorithms used in this work. This research was supported by the Spanish Government Grant No. TINN2011-28753-C02-02.

\newpage
%
%

\end{document}